\documentclass[
reprint,
superscriptaddress,
amsmath,
amssymb,
aps,
prl,
floatfix,
]{revtex4-1}

\usepackage{footnote}	
\usepackage{braket}
\usepackage{dsfont}
\usepackage{chemformula}
\usepackage{subfigure}
\usepackage{graphicx}
\usepackage{dcolumn}
\usepackage{bm}
\usepackage{relsize}
\usepackage{slashed}
\usepackage{dirtytalk}
\usepackage{amsmath}
\usepackage{siunitx}
\usepackage{layouts}
\usepackage{xspace}
\usepackage{graphicx,color,hyperref}
\usepackage{soul} 
\hypersetup{colorlinks=true, linkcolor=blue, citecolor=blue, urlcolor=blue} 
\usepackage{braket}

\begin{document}

\title{Universal Multifractality at the Topological Anderson Insulator Transition}


\author{Ksenija Kovalenka}
\email{ksenija.kovalenka@manchester.ac.uk}
\affiliation{%
Department of Physics and Astronomy, The University of Manchester, Oxford Road, Manchester M13 9PL, United Kingdom
}%


\author{Ahmad Ranjbar}
\affiliation{%
Center for Advanced Systems Understanding, Untermarkt 20, D-02826 Görlitz, Germany
}%
\affiliation{%
Helmholtz-Zentrum Dresden-Rossendorf, Bautzner Landstra\ss e 400, D-01328 Dresden, Germany
}%
\affiliation{%
Dynamics of Condensed Matter and Center for Sustainable Systems Design, Theoretical Chemistry, University of Paderborn, Warburger Str.100, D-33098 Paderborn, Germany
}%

\author{Sam Azadi}
\affiliation{%
Department of Physics and Astronomy, The University of Manchester, Oxford Road, Manchester M13 9PL, United Kingdom
}%
\author{Rodion Vladimirovich Belosludov}
\affiliation{Institute for Materials Research, Tohoku University, Sendai 980-08577, Japan}

\author{Thomas D. K\"{u}hne}
\affiliation{%
Center for Advanced Systems Understanding, Untermarkt 20, D-02826 Görlitz, Germany
}%
\affiliation{%
Helmholtz-Zentrum Dresden-Rossendorf, Bautzner Landstra\ss e 400, D-01328 Dresden, Germany
}%
\affiliation{%
TU Dresden, Institute of Artificial Intelligence, Chair of Computational System Sciences, N\"{o}thnitzer Stra\ss e 46, D-01187 Dresden, Germany
}%

\author{Mohammad Saeed Bahramy}
\email{m.saeed.bahramy@manchester.ac.uk}
\affiliation{%
Department of Physics and Astronomy, The University of Manchester, Oxford Road, Manchester M13 9PL, United Kingdom
}%

\date{\today}

\begin{abstract}
Disorder is ubiquitous in quantum materials, and its interplay with topology can generate phases absent in the clean limit. Using the Haldane model as a minimal setting, we show that disorder not only shifts topological boundaries but also stabilizes a topological Anderson insulator (TAI) between trivial and Chern insulating regimes. Employing the local Chern marker as a real-space topological probe, we map the full phase diagram and demonstrate that the TAI forms a finite domain bounded by trivial and Anderson insulators. Multifractal analysis of low-energy eigenstates at the boundary reveals universal critical spectra, independent of whether disorder generates or destroys topology. These results place topology, localization, and criticality within a unified framework and provide clear benchmarks for real-space diagnostics of disordered topological phases.
\end{abstract}

\maketitle

\begin{figure}
\includegraphics[width=8cm]{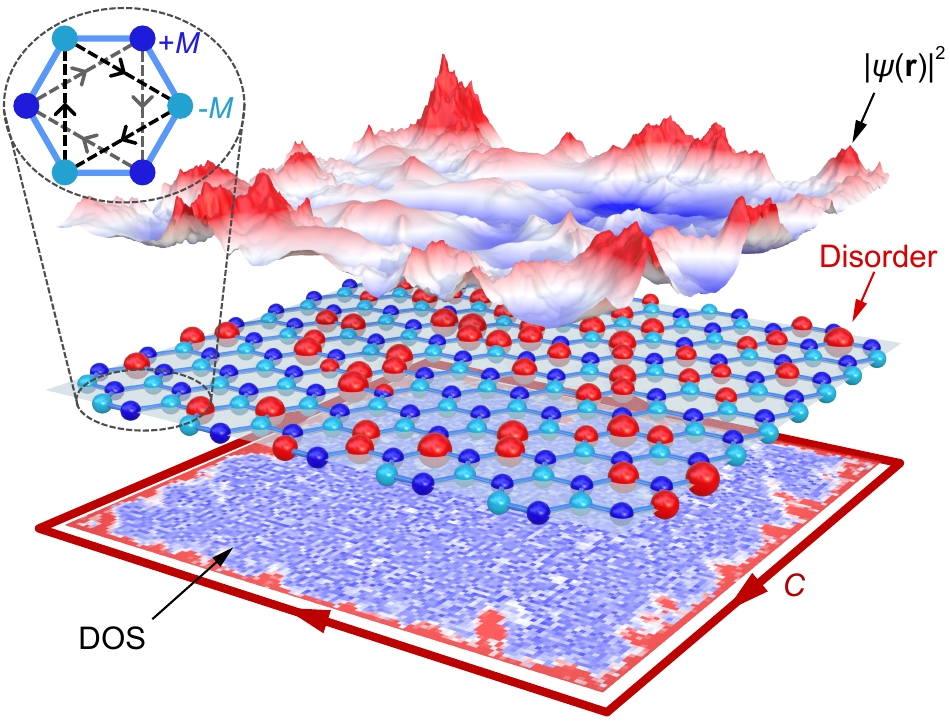}
\caption{\label{fig:fig1} Schematic of the disorder-induced topological Anderson insulator in the Haldane model on a honeycomb lattice. Sublattice staggering $\pm M$ breaks inversion symmetry (blue/cyan sites), while complex next-nearest-neighbour hopping breaks time-reversal symmetry (indicated by arrows in the inset). The upper surface illustrates a representative real-space probability density $|\psi(\mathbf{r})|^2$, and the lower plane shows the (local) density of states at energy $E=0$, highlighting edge-localized spectral weight consistent with a nontrivial Chern phase ($C \neq 0$).}
\end{figure}

The interplay between topology and disorder has emerged as a central theme in condensed matter physics~\cite{PhysRevLett.119.183901,experimentTAI1,experimentTAI2,PRUISKEN1984277,PhysRevB.94.140505,Sau2012}. While disorder is generally expected to localize states and suppress coherent transport, it can also give rise to novel phases of matter~\cite{PhysRevB.97.024204,PhysRevLett.112.016402}. A striking example is the \emph{topological Anderson insulator} (TAI), where moderate disorder converts a trivial band insulator into a topological phase by renormalizing the effective band mass of the wave function\cite{Li2009,Groth2009,Jiang_2009,Guo2010,Meier2018,eblbhz}, as schematically illustrated in Fig.~~\ref{fig:fig1}. This counterintuitive mechanism highlights that disorder does not merely destroy topology but can in fact generate it.

The TAI belongs to a broader set of phenomena in which non-trivial topology arises from the subtle competition between band structure and disorder. In two dimensions, a canonical platform for exploring disorder-driven topological transitions is the Haldane model: a time-reversal–broken honeycomb-lattice Chern insulator in symmetry class A with integer Chern number $C\in\mathbb{Z}$~\cite{Haldane-orig,thonhauser2006insulator}, the same universality class as the integer quantum Hall effect (IQHE)~\cite{PhysRevB.93.245414,Evers2008, KRAMER2005211}. Adding on-site Anderson disorder preserves this classification, yet it induces transitions between topological regimes with distinct Chern numbers (Fig.~\ref{fig:fig1}). These transitions are expected to display universal critical behavior, yet the structure of the critical wave functions at disorder-driven boundaries remains unresolved.

A powerful framework for characterizing how such wave functions evolve at disorder-driven phase boundaries is \emph{multifractal analysis}. 
At criticality, electronic states develop highly inhomogeneous amplitude fluctuations whose statistical properties are encoded in the moments of the probability distribution~\cite{Castellani_1986,Mandelbrot_1974}. 
These moments scale with distinct exponents, giving rise to a nonlinear spectrum of generalized dimensions $D_q$~\cite{MFAnderson,applicationsreview,Halsey1986,Evers2008,ChhabraJensen1989,multifAndtech}, 
where the index $q$ selectively probes different intensity regions of the wave function (large positive $q$ emphasizing rare high-amplitude peaks, small positive $q$ probing broader support). 
Such multifractal spectra are hallmarks of quantum criticality and appear prominently at the IQHE plateau transition~\cite{Huckestein1995}. 
Whether analogous universal spectra also characterize the critical states at disorder-induced topological transitions---such as the boundary of the TAI---remains an open question.

Here we address this question by studying the disordered Haldane model using two complementary tools. First, we construct the phase diagram for topological orders by computing the \emph{local Chern marker} (LCM) in finite open samples \cite{lcm,Prodan2011}. In such systems, since the global sum of the LCM vanishes, a robust imbalance between bulk and edge contributions provides a sensitive indicator of topology even in the absence of translational symmetry. We employ an edge-integrated LCM to cleanly resolve the trivial, topological, and disorder-induced TAI regimes.

Second, we analyze the scaling properties of eigenstates near the Fermi level using finite-size multifractal analysis. We find that along the TAI boundary the generalized-dimension spectra $D_q$ for $q>0$ collapse across widely separated points in the mass-disorder phase diagram, revealing a \emph{universal multifractal fingerprint} of the critical states within this symmetry class. In contrast, deep in the clean topological regime the spectra are edge-dominated, with $D_q\!\to\!1$ for large positive $q$, whereas in the trivial insulating regime the bulk states are localized and $D_q\!\to\!0$. These results demonstrate that multifractality not only provides a sharp diagnostic of the topological phase diagram but also uncovers universal features of criticality in disordered Chern insulators.

We begin by formulating the tight-binding Haldane model for spinless fermions on a honeycomb lattice, defined by the Hamiltonian
\begin{align}
H &= -t_1 \sum_{\langle i,j\rangle} \left(a_i^\dagger a_j + \text{h.c.}\right) 
- t_2 \sum_{\langle\langle i,j\rangle\rangle} \left(e^{i\phi_{ij}} a_i^\dagger a_j + \text{h.c.}\right) \notag \\
&\quad + M \sum_{i\in A} n_i - M \sum_{i\in B} n_i 
+ \sum_i \epsilon_i n_i ,
\label{eq:haldanetb}
\end{align}
where $a_i^\dagger$ and $a_i$ are creation and annihilation operators, $n_i=a_i^\dagger a_i$ is the number operator, and $\langle i,j\rangle$, $\langle\langle i,j\rangle\rangle$ denote nearest- and next-nearest-neighbor pairs, respectively. The nearest-neighbor hopping $t_1$ produces gapless Dirac cones at the Brillouin-zone corners. The complex second-neighbor hopping $t_2 e^{i\phi_{ij}}$ breaks time-reversal symmetry and generates a topological mass gap, while the staggered sublattice potential (mass) $\pm M$ breaks inversion symmetry and competes with the topological mass. The final term represents onsite Anderson disorder with random $\epsilon_i$ drawn from a uniform distribution $\epsilon_i\in[-W/2,W/2]$. All numerical values of the parameters in the subsequent discussion are given in the units of $t_1$.

\begin{figure}
\includegraphics[width=0.45\textwidth]{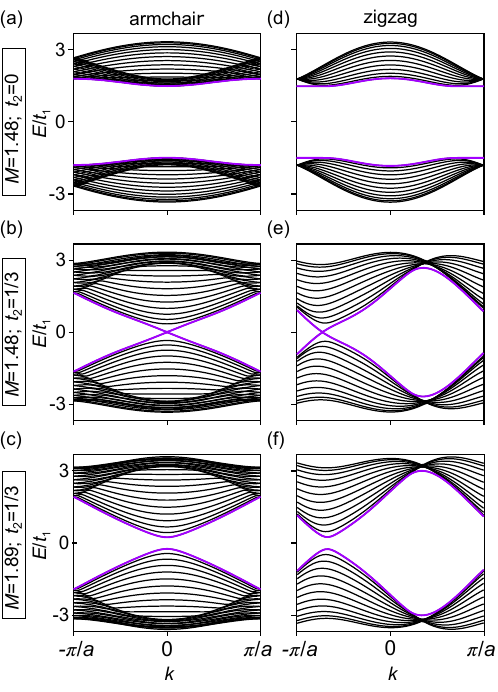}
\caption{\label{fig:Haldane} Energy dispersion of a disorder-free ($W=0$) Haldane nanoribbon with 12 unit cells across. Panels (a)–(c) show the armchair edge for $M=1.48$, $t_{2}=0$ (trivial insulator), $M=1.48$, $t_{2}=1/3$ (Chern insulator), and $M=1.89$, $t_{2}=1/3$ (trivial insulator). Panels (d)–(f) display the corresponding spectra for the zigzag edge. The top valence and bottom conduction bands are highlighted in purple.
}
\end{figure}

The relative strength of mass $M$ and $t_2$ determines whether the clean system is topological or trivial. Figure~\ref{fig:Haldane} shows the band structure of ribbons with open boundaries, demonstrating how the gap character evolves as parameters are tuned. For $M\neq 0$ and $t_2=0$, the system is a trivial band insulator (BI) due to inversion symmetry breaking [Fig.~\ref{fig:Haldane}(a,d)]. Adding a finite $t_2$ at fixed $M$ induces topological edge modes that traverse the bulk gap, characteristic of a Chern insulator (CI) [Fig.~\ref{fig:Haldane}(b,e)]. Increasing mass $M$ further eventually closes the topological gap and restores triviality [Fig.~\ref{fig:Haldane}(c,f)]. This interplay shows that the Haldane model permits tuning between trivial and topological insulating regimes by controlling $M$ and $t_2$.

\begin{figure*}
\includegraphics{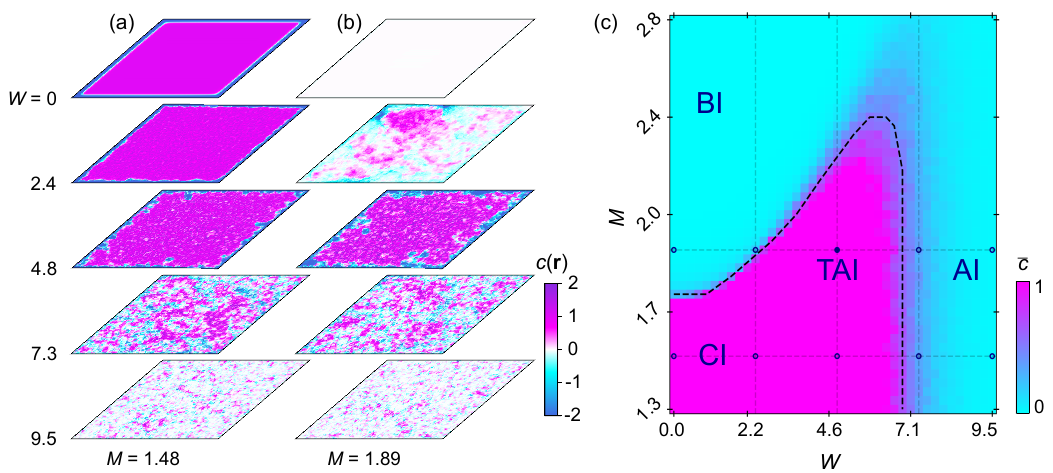}
\caption{\label{fig:chern_h} Disorder dependence of the local Chern marker $c(\mathbf{r})$, computed on a $120\times120$ Haldane sample with $t_{2}=1/3$ for (a) $M=1.48$ and (b) $M=1.89$. The colour map is clipped to $c(\mathbf{r}) = \pm 2$. (c) Corresponding phase diagram as a function of $W$ and $M$, obtained from the integrated LCM $\bar{c}$ along the edges of the finite sample and averaged over 10 disorder realizations. Values of $\bar{c}$ are normalized by the mean across the phase diagram. To emphasize the contrast between trivial ($\bar{c}=0$) and topological ($\bar{c}=1$) regions, the colormap is clipped at $\bar{c}<1$. The guiding line indicates the approximate phase boundary separating the band insulator (BI) and Anderson insulator (AI) from the Chern insulator (CI) and topological Anderson insulator (TAI).
}
\end{figure*}

To explore the role of disorder, we add the onsite Anderson term in Eq.~\eqref{eq:haldanetb}, which breaks translational symmetry and renders momentum-space invariants—such as the Berry curvature or Chern number—ill-defined. Instead, we characterize topology directly in real space using LCM \cite{lcm,Prodan2011,LCMconcept}, a spatially resolved quantity defined for each unit cell located at $\mathbf{r}$ as
\begin{equation}
c(\mathbf{r}) = - \frac{4\pi}{A_c} \textrm{Im}\Big[ \textrm{Tr}_{\textbf{uc}}(\hat{P}\hat{x}\hat{Q}\hat{y})\Big],
\label{eq:locchern}
\end{equation}
where the trace is taken over a unit cell of area $A_c$. $\hat P$ is the projector onto occupied states, $\hat Q=1-\hat P$, and $\hat x,\hat y$ are position operators. In periodic systems, the average of $c(\mathbf{r})$ converges to the bulk Chern number. In finite open systems, however, the sum satisfies $\sum_r c(\mathbf{r})=0$, and nontrivial topology manifests as a robust imbalance between bulk and edge contributions. We exploit this property by integrating $c(\mathbf{r})$ over a three-row boundary strip to define the edge-integrated marker $\bar c$, normalized by the global average across the $[M, W]$ phase diagram. A finite $\bar c$ thus signals a topological phase even in the presence of disorder.

Figure~\ref{fig:chern_h} (a,b) illustrates how the local Chern marker captures the disorder-driven evolution of topology in the Haldane model for the mass values used in Fig.~\ref{fig:Haldane}. As shown in Fig.~\ref{fig:chern_h}(a), for the CI phase ($M=1.48$), in the clean limit ($W=0$) the system exhibits a uniform positive bulk contribution of $c(\mathbf{r})$ compensated by opposite edge modes, as expected for a Chern insulator and consistent with the requirement $\sum_r c(\mathbf{r})=0$ under open boundary conditions. Introducing disorder, the bulk coherence of $c(\mathbf{r})$ is progressively lost and eventually replaced by short-range fluctuations, marking the transition to a trivial Anderson insulator (AI).

In contrast, the initially trivial BI phase ($M=1.89$) undergoes a qualitatively different evolution as can be seen in Fig.~\ref{fig:chern_h}(b). At weak disorder, $c(\mathbf{r})$ is featureless and localized near zero, reflecting triviality. With increasing $W$, however, an extended bulk pattern of $c(\mathbf{r})$ emerges, accompanied by compensating edge signals—clear evidence that disorder has induced a topological phase. This is the hallmark of the TAI. Upon further increasing $W$, the bulk signal again breaks down into random short-range fluctuations, indicating that the system enters a trivial Anderson insulating regime. The LCM thus directly visualizes the two-step evolution of the trivial system: {BI $\rightarrow$ TAI $\rightarrow$ AI}.

\begin{figure*}
\includegraphics[width=\textwidth]{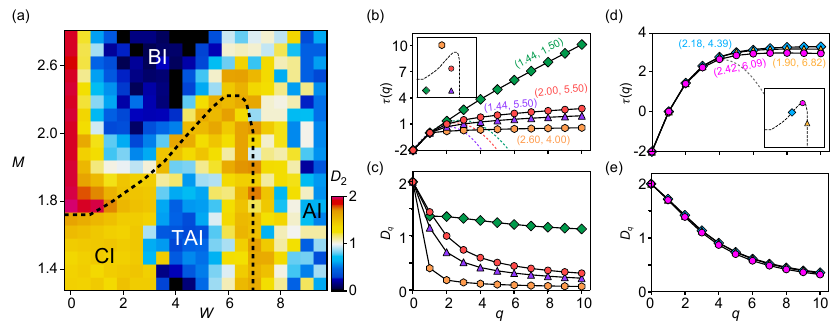}
\caption{\label{fig:multifractalify} 
(a) Map of the correlation dimension $D_{2}$ of Haldane-model eigenstates in the $(M,W)$ plane. 
The dashed curve indicates the disorder-driven phase boundary extracted from the LCM analysis in Fig.~\ref{fig:chern_h}(c). 
(b,c) Disorder-averaged multifractal spectra $\tau(q)$ and generalized dimensions $D_{q}$ evaluated at representative points in the BI (orange hexagons), CI (green diamonds), and TAI (red circles and purple triangles) phases. 
Dashed curves show best fits to the parabolic approximation, Eq.~\ref{eq:WZNW}. 
(d,e) Same as (b,c), but for three representative parameter sets along the phase boundary, demonstrating the collapse of $\tau(q)$ and $D_q$ onto universal critical curves. In (b-e), the errorbars are included but due to their insignificance appear invisible and hidden behind the markers.
}
\end{figure*}

Figure~\ref{fig:chern_h}(c) summarizes these behaviors across the $(M,W)$ phase diagram using the edge-integrated marker $\bar c$, averaged over ten disorder realizations at each point. The color scale is limited to $\bar c\le1$ for clarity, although values in the clean topological regime can be substantially larger. Two features stand out. First, the topological domain ($\bar{c} \ge 1$) forms a contiguous topological phase region that connects the clean CI region at small $M$ with the disorder-stabilized TAI region at intermediate $W$. Its low-disorder boundary follows the clean band inversion, while the intermediate-disorder boundary reflects the renormalization of the Dirac mass that drives a trivial system into the topological sector. Second, this region is bounded on both sides by insulating phases, where $\bar c\!\approx\!0$: on the large-$M$ side by the trivial BI phase, and on the large-$W$ side by the AI phase, where strong disorder destroys the extended $c(\mathbf{r})$ patterns and suppresses $\bar c$. In physical terms, moderate disorder opens a mobility window that sustains chiral boundary channels, but once scattering becomes strong enough, the coherence required for a finite LCM is lost. The $\bar c \sim 1$ contour (guiding line) thus encloses a topological island in parameter space, surrounded by topologically trivial BI and AI regimes.

To probe the critical states across the phase diagram in Fig.~\ref{fig:chern_h}, we use a finite-size multifractal analysis based on box-counting \cite{Halsey1986,Evers2008,ChhabraJensen1989}. We partition an $L\times L$ sample into $N_l=(L/l)^2$ non-overlapping boxes of side $l$ and define the box probabilities based on the wavefunction $\psi$ at $i$th sites in the box
\begin{equation}
\mu_k(l)=\sum_{i\in k\;\mathrm{th\;box}} |\psi_i|^2,\qquad \sum_{k=1}^{N_l}\mu_k(l)=1.
\end{equation}
The generalized inverse participation ratios (IPRs) are then
\begin{equation}
P_q(l)=\sum_{k=1}^{N_l} \mu_k(l)^q,
\end{equation}
which emphasize high-amplitude regions for $q>0$ and low-amplitude tails for $q<0$ (cf. Fig.~\ref{fig:fig1}(c)). The resulting IPRs scale as,
\begin{equation}
P_q(l)\propto \lambda^{\tau(q)},\qquad \lambda=\frac{l}{L},\qquad \tau(q)=(1-q)D_q,
\end{equation}
where the scaling exponent $\tau(q)$ is related to the generalised dimension $D_q$. Consequently, $D_q$ for each of the moments can be expressed as 
\begin{equation}
D_q=\frac{1}{1-q}\,\lim_{\lambda\to 0}\frac{\ln P_q(\lambda)}{\ln \lambda}.
\end{equation}
For extended states, the generalized dimension equals the dimension of the support of the measure, $D_q=d$ ($d=2$ here). Exponentially localized states give $D_q=0$ for $q>0$, while critical states display a nonlinear $q$ dependence, the hallmark of multifractality. The case $q=1$ is equivalent to the Shannon entropy \footnote{For $q=1$, one uses $D_1=-\lim_{\lambda\to 0}\sum_k \mu_k(l)\ln\mu_k(l)/\ln \lambda$ \cite{Halsey1986}.}. Generalised dimensions for $q<0$ reflect the scaling of the low-intensity regions of the wavefunction, which makes them particularly sensitive to numerical precision. For this reason, we omit them from our analysis.

Numerically, we target the four eigenstates closest to $E=0$ using the implicitly restarted Arnoldi method \footnote{The Arnoldi method is implemented with convergence tolerance $\|H\psi-\lambda\psi\|\leq 10^{-3}$.} ~\cite{arpackdoc}, and extract the exponents from ensemble averages of $P_q$ over disorder realizations \footnote{The exponents are obtained by a linear fit through $log(\langle P_q(\lambda)\rangle )$ against $log( \lambda)$ for $l$ = {10, 12, 15, 20, 24, 30, 40, 60} and $L$ = 120. Here $\langle P_q \rangle$ denotes an arithmetic average.}, which, contrary to typical averaging, treats all disorder realisations on equal grounds \cite{VasquezEnsemble, PhysRevB.62.7920}. A representative map of $D_2$ across the $(M,W)$ plane is shown in Fig.~\ref{fig:multifractalify}(a); the guiding line coincides with the phase boundaries from Fig.~\ref{fig:chern_h}(c), enabling a direct comparison between trivial, topological, and boundary-critical regimes.

As is evident from Fig.~\ref{fig:multifractalify}(a), in the trivial BI region the eigenstates are exponentially localized, leading to a vanishing correlation dimension $D_{2}\simeq 0$. 
In contrast, deep in the clean CI phase the statistics are strongly influenced by the presence of chiral boundary modes. 
As a result, the wave-function intensity acquires a characteristic nonzero fractal dimension, $D_{2}\simeq 1.5$, reflecting the pronounced edge contribution. 
This edge--bulk imbalance provides a distinct multifractal fingerprint of the CI regime, sharply differentiating it from the adjacent disorder-induced TAI, where the bulk becomes progressively localized and $D_{2}$ is again suppressed.

Most remarkably, Fig.~\ref{fig:multifractalify}(a) reveals an extended ridge of nearly identical $D_{2}$ values tracing the disorder-driven phase boundary previously identified through the LCM analysis. 
This feature persists irrespective of whether the transition is approached from the trivial side, where disorder induces critical bulk weight, or from the topological side, where disorder suppresses coherent edge transport. 
The emergence of this continuous critical manifold strongly suggests that the topological transition is governed by universal multifractal scaling rather than phase-specific microscopic details.

\begin{table}[t]
\caption{Extracted anomalous multifractal exponents $\gamma$ obtained by fitting the numerical $\tau(q)$ spectra in the different phases to the parabolic approximation, Eq.~\ref{eq:WZNW}.
}
\begin{ruledtabular}
    \begin{tabular}{ll|l}
    Phase &$(M,W)$ & $\gamma$ \\
    \hline  
       CI  & $(1.44,1.5)$ & 0.321\\
       TAI & $(2.00,5.50)$ &0.500\\
       TAI & $(1.44,5.50)$ &0.650\\
       Phase Boundary & $(2.18,4.39)$ & 0.277
    \end{tabular}
    
    \label{tab:1}
    \end{ruledtabular}
\end{table}
To gain deeper insight into the multifractal structure of the disordered eigenstates, we present in Fig.~\ref{fig:multifractalify}(b--e) the $q$-dependence of the scaling exponents $\tau(q)$ and the generalized fractal dimensions $D_q$, averaged over 2000 disorder realizations at representative points in the $(M,W)$ plane. 
Across the different phases, $\tau(q)$ exhibits a pronounced nonlinearity in $q$ [Figs.~\ref{fig:multifractalify}(b,c)], demonstrating that the wave functions are neither purely localized nor fully ergodic, but instead display genuine multifractal correlations.

Focusing on $\tau(q)$, one observes for each parameter set an initial nonlinear regime followed by an approximately linear large-$q$ behavior. 
The nonlinear sector captures the range of moments for which multifractal fluctuations are most prominent and can be effectively described within the Wess--Zumino--Novikov--Witten framework~\cite{PhysRevB.75.184201}. 
A widely used fit (often termed the parabolic approximation) takes the form~\cite{PhysRevLett.101.116803,recentrevAnd}
\begin{equation}
    \tau(q)=d(q-1)+\gamma\,q(1-q),
    \label{eq:WZNW}
\end{equation}
where $d=2$ is the spatial dimension and $\gamma$ quantifies the anomalous multifractal contribution. 
The dashed curves in Fig.~\ref{fig:multifractalify}(b) show the best parabolic fits in each phase, and the extracted $\gamma$ values are summarized in Table~\ref{tab:1}. 

In the CI phase, the parabolic regime extends over the widest interval of $q$, corresponding to the smallest effective $\gamma$~\cite{recentrevAnd}, consistent with multifractality predominantly governed by edge-dominated critical states~\cite{PhysRevLett.101.116803}. 
At larger $q$, deviations from parabolicity become apparent, with $\tau(q)$ crossing over into a monotonic behavior reminiscent of critical spectra in power-law random banded matrix models~\cite{PhysRevB.62.7920}. 
In the TAI regime, by contrast, the parabolic window is significantly reduced and followed by a plateau-like tendency, signaling a progressive \emph{termination} of multifractality~\cite{Evers2008}. 
This reflects the onset of disorder-induced localization, as also evident in Fig.~\ref{fig:multifractalify}(c), where $D_q$ flows toward the BI-like limit $D_q\to 0$ at large $q$.

Most strikingly, at the disorder-driven phase boundary both $\tau(q)$ and $D_q$ collapse onto nearly universal curves, independent of the particular choice of $(M,W)$ along the transition line [Figs.~\ref{fig:multifractalify}(d,e)]. 
In this critical regime, the parabolic approximation remains valid over an even broader range of $q$ than in the CI phase, yielding an anomalous exponent as small as $\gamma\simeq 0.277$ (Table~\ref{tab:1}). 
Remarkably, this value is close to $\gamma=0.262$ reported for the integer quantum Hall plateau transition~\cite{PhysRevB.64.241303}. 
Consistently, the extracted correlation dimension $D_{2}\simeq 1.40$ from Fig.~\ref{fig:multifractalify}(c) lies near the established IQHE critical value $D_{2}\simeq 1.52$~\cite{PhysRevB.59.9714}. 

These results indicate that the disorder-driven topological transition in the Haldane model belongs to the same unitary universality class as the quantum Hall critical point, thereby linking lattice Chern insulators to the paradigmatic multifractal criticality of the IQHE. 
Together with the LCM phase diagram, our findings demonstrate that the disordered Haldane model hosts a robust topological region---including both the clean CI and the disorder-induced TAI---surrounded by trivial and Anderson insulating phases, with all intervening transitions governed by universal multifractal scaling.

In summary, we have demonstrated that disorder fundamentally reshapes the topological phase diagram of the Haldane model. Using the real-space local Chern marker, we established the emergence of a disorder-stabilized topological region, including TAI, and mapped its boundaries against trivial and Anderson insulating regimes. Multifractal analysis of critical wave functions revealed that these boundaries are governed by universal scaling, independent of whether topology is created or destroyed by disorder. Together, these results show that the disordered Haldane model provides a minimal platform in which topology, localization, and criticality can be understood within a unified framework. Beyond this specific model, our findings highlight general mechanisms by which disorder can both stabilize and destabilize topological phases, offering testable predictions for transport and spectroscopic probes in engineered lattices and correlated materials.


\begin{acknowledgments}
We acknowledge the support of the Leverhulme Trust under the grant agreement RPG-2023-253.  We also thank the Paderborn Center for Parallel Computing (PC2) for allocations on thier high-performance computational facilities (Noctua2) as well as the Center for Computational Materials Science at Institute for Materials Research (IMR) for allocations on the MASAMUNE-IMR supercomputer system (Project No. 202112-SCKXX-0510). K.K. wishes to acknowledge the financial support of the CDT in Graphene NOWNANO.
\end{acknowledgments}









\bibliography{references}

\end{document}